\documentstyle[multicol,eqsecnum,aps,epsf,prl]{revtex}

\newcommand \comma {\mbox{\makebox[.1 in]{ },}}
\newcommand \perd {\mbox{\makebox[.1 in]{ }.}}
\newcommand \gr {|g\rangle}

\newcommand \doe {|D\rangle}

\newcommand \ace {|A\rangle}

\newcommand \od { a_{_D}}
\newcommand \odd {a^\dagger_{_D}}
\newcommand \oa {a_{_A}}
\newcommand \oda {a^\dagger_{_A}} 
\newcommand \den {\epsilon_{_D}}
\newcommand \aen {\epsilon_{_A}}
\newcommand \tden {\tilde \epsilon_{_D}}
\newcommand \taen {\tilde \epsilon_{_A}}
\newcommand \ocd {\odd\od}
\newcommand \oca {\oda\oa}
\newcommand \omn {\omega_n}
\newcommand \bn {b_n}
\newcommand \bdn {b^\dagger_n}
\newcommand \ocb {\bdn\bn}
\newcommand \gnd {g_{nD}}
\newcommand \gna {g_{nA}}
\newcommand \gndt {g_{n\Delta}}

\newcommand \thdd {\theta_D^\dagger}
\newcommand \thd {\theta_D}
\newcommand \thda {\theta_A^\dagger}
\newcommand \tha {\theta_A}
\newcommand \delte {\delta \tilde \epsilon}

\newcommand \btha {B_{_A}}
\newcommand \bthd {B_{_D}}

\def \be {\begin{equation}}
\def \ee {\end{equation}}
\def \ben {\begin{eqnarray}}
\def \een {\end{eqnarray}}

\begin{document}
\bibliographystyle{/afs/athena.mit.edu/user/y/o/younjoon/paper/bibtex/aip}
\draft

\title{Nonequilibrium generalization of F\"{o}rster-Dexter theory for 
excitation energy transfer\footnote{published in {\it Chemical Physics}, 
{\bf 275}, pp.~319--332 (2002)}}

\author{Seogjoo Jang, YounJoon Jung, and Robert J. Silbey}
\address{Department of Chemistry, Massachusetts Institute of Technology, Cambridge, Massachusetts 02139} 
\date{\today}
\maketitle

\begin{abstract}
F\"{o}rster-Dexter theory for excitation energy transfer is generalized
for the account of short time nonequilibrium kinetics due to the 
nonstationary bath relaxation.  The final rate expression is presented 
as a spectral overlap between the time dependent stimulated 
emission and the stationary absorption profiles, which allows 
experimental determination of the time dependent rate.  For a harmonic 
oscillator bath model, an explicit rate expression is derived and 
model calculations are performed in order to examine the dependence
of the nonequilibrium kinetics on the excitation-bath coupling strength 
and the temperature.   Relevance of the present theory with recent 
experimental findings and possible future theoretical directions 
are discussed.

\end{abstract}

\begin{multicols}{2}

\vskip 0.5in
\section{Introduction}
Excitation energy transfer (EET)\cite{forster-dfs,forster-book,dexter-jcp,agranovich} is ubiquitous in photo sensitive materials\cite{eet-org,birks,ret,cerullo-cpl335,meskers-cpl339,buckley-cpl339,neuwahl-jpcb105,Morgado-jmc11} and is one of the key steps in photosynthesis\cite{ret,hu-pt,sundstrom-jpcb103,renger-pr343}.  
Since the seminal works of F\"{o}rster\cite{forster-dfs,forster-book} 
and Dexter (FD)\cite{dexter-jcp}, their rate expressions have 
been confirmed by numerous experiments and have played fundamental 
roles in understanding various luminescence phenomena\cite{agranovich,eet-org,birks,ret,cerullo-cpl335,meskers-cpl339,buckley-cpl339,neuwahl-jpcb105,Morgado-jmc11}.  
In these advances, the spectral overlap 
expression of F\"{o}rster\cite{forster-dfs,forster-book}, 
which allows identification of the reaction rate without any resort 
to a model Hamiltonian, has been essential.  

The rate expressions of FD are applications of the Fermi's golden rule (FGR),
which rely on the smallness of the resonance interactions.   
The spectral overlap expression of F\"{o}rster derives from the additional 
simplification that bath modes coupled to the energy donor and 
acceptor are independent of each other.  If there are {\it common}
bath modes\cite{soules-prb3}, such a simple expression is not in general valid.  
However, within the harmonic oscillator bath model, 
rigorous extensions of the FD theory can be 
made\cite{soules-prb3,rackovsky-mp25,jackson-jcp78,silbey-arpc27} with 
the use of small polaron transformation, and can be further generalized 
for the study of long-time dynamics based on the Redfield-type 
equation\cite{rackovsky-mp25} and for the understanding of  
exciton transport in molecular crystals\cite{silbey-arpc27,grover-jcp54,munn-jcp68}. 
In recent years, theoretical advances to cope with new 
experiments have been made, such as microscopic 
consideration of the medium effect\cite{juzeliunas,hsu-jcp114}, generalization of 
the FD rate expression for disordered multichromorphic 
systems\cite{sumi-jpcb103,mukai-jpcb-103,mukai-jl87-89,scholes-jpcb104,scholes-jpcb105,damjanovic-pre59,yeow-jpca104}, and unified theories covering up to the 
intermediate and strong coupling regimes\cite{lin-pre,kakitani-jpcb103,kimura-jl87-9,kimura-jpcb104}.

Although often not noticed, the assumption of stationarity is 
implicit in the FD theory.  The FGR is valid only when the initial
density operator commutes with the zeroth order Hamiltonian and the
bath time scale is much shorter than that of the electronic transition. 
For EET processes occurring in nanosecond or longer time scale, 
one may safely assume that the bath modes have already relaxed and 
become equilibrated with the excited donor before the energy transfer 
takes place, unless there are ultra-slow modes or spin interactions 
with comparable time scales.  Application of the FGR, thus the FD 
theory, can be justified for this case.  However, EET in general 
is a nonstationary process where nonequilibrium relaxation of the
nuclear degrees of freedom occurs during and after the electronic
excitation.  For fast EET processes occurring in a time scale 
comparable to the bath relaxation time, the reaction kinetics 
predicted by the FD theory may not be accurate enough.  

The importance of the nonequilibrium effects for fast EET was 
in fact recognized long time ago, and was named as hot transfer\cite{tekhver-jetp42,ht,sumi-jpsj51,sumi-prl50}.
Due to experimental limitations, however, the earlier experiments
and relevant theories were concerned with the frequency domain situation 
where the reaction rate at issue is the stationary long time limit 
in the presence of an excitation field tuned for the hot transfer\cite{tekhver-jetp42,ht}.
Time dependent pump-probe situation was considered later by Sumi\cite{sumi-jpsj51,sumi-prl50}, who formulated a time dependent generalization of the 
FD theory based on a nonequilibrium golden rule 
approximation\cite{sumi-jpsj51}.

With the advance of ultrafast spectroscopy, it has become possible 
to induce the electronic excitation in the femtosecond scale.  
Experiments\cite{cerullo-cpl335,neuwahl-jpcb105,sundstrom-jpcb103,gnanakaran,vangrondelle,king-jpcb105} performed in this manner reveal systems where the EET 
rate is comparable to the vibrational relaxation rate of some modes.  
A typical example for this is the EET from B800 to B850 in the 
light harvesting complex 2 (LH2)\cite{sundstrom-jpcb103}, which is known 
to occur in about $1 \ {\rm ps}$.   Evidences for fast EET were found in other 
systems also such as the conjugated polymer\cite{cerullo-cpl335}, dendrimers\cite{neuwahl-jpcb105}, and the 
photosynthetic reaction center\cite{king-jpcb105}.
In general, the microscopic mechanisms of the EET in these systems 
are quite complicated, and considerations of 
multipolar transitions, multichromorphic effects, and disorder may 
be important.  However, simple comparison of time scales indicates
that examination of the nonequilibrium effects cannot be overlooked.
The theory of Sumi\cite{sumi-jpsj51,sumi-prl50} 
and recent theories\cite{kakitani-jpcb103,kimura-jl87-9,kimura-jpcb104} 
on intermediate and strong 
coupling seem suitable for these considerations.  However, the nature
of the approximation involved in the bath relaxation dynamics, upon
which the nonequilibrium kinetics is sensitive, is not clear in these
approaches.

In the present work, we provide a straightforward and rigorous 
nonequilibrium generalization of the FD theory.  The procedure 
is to go through the same perturbation theory as in deriving the FGR, 
but starting from the nonstationary initial states and considering 
full time dependences.  Our analysis is limited to the usual perturbation
regime of the FD theory.  We assume that the resonance interaction 
is small enough to validate second order perturbation theory and that the 
reaction is virtually irreversible due to either energetic or entropic 
reason. However, the excitation bath coupling is treated rigorously.   
The present theory is analogous to the nonequilibrium generalization 
of the electron transfer\cite{cho-jcp103}, but an important distinction 
is that we provide a spectral overlap expression valid for 
arbitrary bath without common modes, a typical situation for the EET.

Our theory can be considered as the pump-probe version of the hot transfer 
rate theories\cite{tekhver-jetp42,ht}.  More importantly, 
our spectral overlap expression brings a connection between the time 
dependent reaction rate and the modern 
ultrafast spectroscopy experiment, which allows direct determination 
of the reaction rate or enables experimental confirmation whether 
the usual assumption of the EET kinetics is valid.   
In addition, we provide calculations of the time dependent rate 
for the model of the harmonic oscillator bath, which illustrate some 
features of the nonequilibrium EET kinetics.

The sections are organized as follows.  In Sec. IIA, we present the
main formalism and provide the spectral overlap expression valid 
for general bath Hamiltonian without common modes.   Section IIB 
provides a complementary result of an explicit rate expression
for the harmonic oscillator bath.  In Sec. III, model calculations
are made.  Sec. IV concludes with summaries and the relevance of 
the present theory with recent experiments.

\section{Theory}
\subsection{General bath without common modes}
The system consists of two distinctive chromophores, 
donor ($D$) and acceptor ($A$).  The state where both $D$
and $A$ are in their ground electronic states is denoted 
as $\gr$.   The state where $D$  is excited 
while $A$ remains in the ground state is denoted 
as $\doe=\odd\gr$, where $\odd$ is the corresponding 
creation operator, and $\ace=\oda\gr$ is defined similarly.  
We assume that both chromophores are excited to singlet 
states and do not consider multiexciton states.  Therefore,
the three electronic states $\gr$, $\doe$, and $\ace$ form a 
complete set of electronic states.  
All the rest of the dynamic degrees of freedom such as molecular 
vibrations and solvation coordinates are included in the bath.  
The bath Hamiltonian is defined as that in the ground state $\gr$ 
and is assumed to be $H_b=H_{b_D}+H_{b_A}$, where the subscripts 
of $D$ and $A$ denote the components coupled to chromophores 
$D$ and $A$, respectively.  
That is, in the present work, the effect of common modes is disregarded.
For EET processes in a medium where phonons and vibrations are localized, 
this assumption seems reasonable.

Since we consider the excitation dynamics during a time much
shorter than the lifetime of the excited state, we neglect 
the spontaneous decay channels of the excited states.  Adopting
a second quantization notation where $\gr$ is treated as if the
vacuum state, the zeroth order Hamiltonian describing the 
interaction-free dynamics can be written as
\be
H_0=\den a_{_D}^{\dagger}a_{_D}+\aen a_{_A}^{\dagger}a_{_A}+H_{b}\comma\label{eq:h0}
\ee
where $\den$ and $\aen$ are excitation energies of $D$ and $A$.

The resonance interaction between $\doe$ and $\ace$ is represented by
\be
H_{DA}=J(\odd\oa+\oda\od)  \comma \label{eq:hda}
\ee
where $J$ is a function of the position vectors and the transition 
dipoles of $D$ and $A$.  The functional form of $J$ depends 
on the mechanism of EET.  For dipole-dipole interaction, it varies
as the inverse third power of the distance between $D$ and $A$.
For exchange interaction, it is an exponential function of the distance.
Here we do not specify the detailed mechanism, but simply assume 
that it is small enough to warrant a perturbation analysis and 
does not have any dependence on the bath operators (vibrational 
degrees of freedom), so called Condon approximation.

The excitation bath coupling is assumed to be as follows:
\be
H_{eb}=B_{_D} a_{_D}^{\dagger}a_{_D}+B_{_A} a_{_A}^{\dagger}a_{_A} \comma\label{eq:heb}
\ee
where $B_{_D}$ and $B_{_A}$ are bath operators coupled to 
$\doe$ and $\ace$ respectively.  These operators and the bath Hamiltonian 
can be arbitrary except for the following condition:   
\be
[H_{b_D},H_{b_A}]=[H_{b_D},B_{_A}]=[H_{b_A},B_{_D}]=0 \comma  \label{eq:nocom2}
\ee
which implies that the bath modes coupled to $\doe$ and $\ace$ are independent 
of each other.  This assumption can be justified if the chromophores 
$D$ and $A$ are far enough apart from each other and the major nuclear modes 
coupled to excitations are localized to either $D$ or $A$, which can be 
consistent with the assumption of smallness of $J$. 

Summing up Eqs. (\ref{eq:h0})-(\ref{eq:heb}), the total Hamiltonian 
describing the dynamics of the chromophores, in the single excitation 
manifold,  and the bath is given by 
\be
H=H_0+H_{DA}+H_{eb}\perd \label{eq:tot-h}
\ee
For $t <0$, the chromophores are assumed to be in the state $\gr$
and thus the total Hamiltonian is equal to $H_b$.   The bath is assumed 
to be in the canonical equilibrium of $H_b$ during this period.  
At time zero, an impulsive pulse  selectively excites $D$.  
Here we approximate this as a delta pulse in time.  
Then, the density operator 
at time zero, right after the irradiation, is given by 
\be
\rho (0)=|D\rangle \langle D|e^{-\beta H_b}/Z_b \comma \label{eq:inic}
\ee
where $\beta=1/k_{_B}T$ and $Z_b=Tr_b\{e^{-\beta H_b}\}$.

The probability at time $t$ for the excitation to be found at 
$A$ is given by 
\be
P_A(t)=Tr_b\left\{\langle A|e^{-iHt/\hbar}\rho (0)e^{iHt/\hbar}|A\rangle \right\} \perd \label{eq:pat0}
\ee
For short enough time compared to $\hbar/J$, a perturbation expansion
of this with respect to $H_{DA}$ can be made.  Inserting the first order 
approximation of the time evolution operator $e^{-iHt/\hbar}$ and the
complex conjugate into Eq. (\ref{eq:pat0}), we obtain the following 
expression valid up to the second order of $J$:   
\ben
P_{A}(t)&\approx& \frac{J^2}{\hbar^2} \int_0^t dt' \int_0^t dt'' e^{i(\epsilon_A-\epsilon_D)(t'-t'')/\hbar} \nonumber \\
&&\times\frac{1}{Z_b}Tr_b\left\{e^{i(H_{b}+\btha)(t'-t'')/\hbar} e^{-i(H_b+\bthd)t'/\hbar}\right .\nonumber \\
&&\times \left.  e^{-\beta H_b} e^{i(H_b+\bthd)t''/\hbar}\right\} \perd
\een
The time dependent EET rate is then defined as the 
derivative of this acceptor probability as follows:
\ben
k(t)&\equiv& \frac{d}{dt}P_{A}(t) \nonumber \\
&\approx&\frac{2 J^2}{\hbar^2} {\rm Re} \left [\int_0^t dt' e^{i(\epsilon_{_D}-\epsilon_{_A})t'/\hbar}\frac{1}{Z_{b}}Tr_{b}\left\{e^{i(H_{b}+\bthd)t/\hbar}\right .\right . \nonumber \\
&&\mbox{\makebox[.4 in] { }}\times \left . \left . e^{-i(H_{b}+\btha)t'/\hbar}e^{-i(H_b+\bthd)(t-t')}e^{-\beta H_{b}}\right\}\right ]  \perd 
\nonumber \\ \label{eq:kt1}
\een
Under the assumption of Eq. (\ref{eq:nocom2}), the trace over 
the bath degrees of freedom comprising $H_{b_D}$ and $H_{b_A}$ 
can be decoupled from each other in the following way:
\ben
k(t)&=&\frac{2J^2}{\hbar^2} {\rm Re} \left [\int_0^t dt' e^{i(\epsilon_{_D}-\epsilon_{_A})t'/\hbar}\right . \nonumber \\
&& \times \frac{1}{Z_{b_A}}Tr_{b_A}\left\{e^{iH_{b_A}t'/\hbar}e^{-i(H_{b_A}+\btha)t'/\hbar}e^{-\beta H_{b_A}}\right\}\nonumber \\
&&\times \frac{1}{Z_{b_D}}Tr_{b_D}\left\{e^{i(H_{b_D}+\bthd)t/\hbar}e^{-iH_{b_D}t'/\hbar}\right . \nonumber \\
&&\mbox{\makebox[.5 in]{ }}\times\left. \left .e^{-i(H_{b_D}+\bthd)(t-t')/\hbar} e^{-\beta H_{b_D}}\right\} \right ]\comma\label{eq:kt2}
\een
where $Z_{b_A}=Tr_{b_A}\{ e^{-\beta H_{b_A}}\}$ and $Z_{b_D}=Tr_{b_D}\{ e^{-\beta H_{b_D}}\}$.  Equation (\ref{eq:kt2}) is the nonequilibrium generalization 
of the FD rate, but expressed in the time domain.  As has been outlined in 
the introduction, this result has been obtained by applying second order 
perturbation theory to the time dependent density operator with the 
nonstationary initial condition of Eq. (\ref{eq:inic}).  
The decoupled form of Eq. (\ref{eq:kt2}) makes it possible to express 
the reaction rate as an overlap of frequency domain 
spectral profiles of independent donor and acceptor, as will be shown later.
Before going through this procedure, it is meaningful to clarify the 
difference between the present result and the FD rate expression.  
Two additional approximations are necessary.  

The first is the 
assumption of stationarity, equivalent to the following replacement: 
\ben
&&e^{-i(H_{b_D}+\bthd)(t-t')/\hbar}e^{-\beta H_{b_D}}e^{i(H_{b_D}+\bthd)(t-t')/\hbar}\nonumber \\
&&\mbox{\makebox[1 in]{ }}\rightarrow \frac{Z_{b_D}}{Z_{b_D}'} e^{-\beta (H_{b_D}+\bthd)} \comma \label{eq:relax}
\een 
where $Z_{b_D}'=Tr_{b_D}\{e^{-\beta (H_{b_D}+\bthd)}\}$.  This approximation 
implies that the nuclear dynamics on the excited donor potential energy 
surface is ergodic in a time scale shorter than that of the EET transfer 
kinetics.  
We define the reaction rate based on this fully relaxed density operator 
as  
\ben
k_r(t)&=&\frac{2J^2}{\hbar^2} {\rm Re} \left [\int_0^t dt' e^{i(\epsilon_{_D}-\epsilon_{_A})t'/\hbar}\right . \nonumber \\
&& \times \frac{1}{Z_{bA}}Tr_{bA}\left\{e^{iH_{bA}t'/\hbar}e^{-i(H_{bA}+\btha)t'/\hbar}e^{-\beta H_{bA}}\right\}\nonumber \\
&&\times \frac{1}{Z_{bD}'}Tr_{bD}\left\{e^{i(H_{bD}+\bthd)t'/\hbar}e^{-iH_{bD}t'/\hbar}\right . \nonumber \\
&&\mbox{\makebox[1 in]{ }}\times\left. \left .e^{-\beta (H_{bD}+B_D)}\right\} \right ]\perd \label{eq:krt-1}
\een
The derivation of this expression can be made by rewriting  $e^{i(H_{bD}+B_D)t/\hbar}$ 
in Eq. (\ref{eq:kt2}) as  $e^{i(H_{bD}+B_D)(t-t')/\hbar}e^{i(H_{bD}+B_D)t'/\hbar}$, 
using the cyclic symmetry of the trace operation for the first term, and then finally 
imposing the replacement of Eq. (\ref{eq:relax}). 

The second is the infinite time approximation.  That is, the FD rate 
corresponds to the following limit: 
\be
k_{_{FD}}=k_r(\infty) \perd
\ee
This approximation is valid if there is time scale separation between 
the bath dynamics and the EET kinetics.

For practical purposes of evaluating the reaction rate for 
real systems, it is important to find the explicit expression
for Eq. (\ref{eq:kt2}) in the frequency domain as an overlap 
of independent spectral profiles of the donor and of the acceptor.
Considering the fact that the term involving the acceptor is 
identical in Eqs. (\ref{eq:kt2}) and (\ref{eq:krt-1}), 
one can expect that the nonequilibrium EET rate also involves
the stationary absorption profile of the acceptor.  For this purpose,
we define the absorption profile of the acceptor as
\ben
I_{A}(\omega)&=&|\mbox{\boldmath $\mu_A$}\cdot {\bf \hat e} |^2 \int_{-\infty}^{\infty} dt\ e^{i\omega t -i\aen t/\hbar}\frac{1}{Z_{bA}} Tr_{bA}\left\{e^{iH_{bA}t/\hbar} \right . \nonumber \\
&&\mbox{\makebox[.8 in]{ }}\times \left .e^{-i(H_{bA}+\btha)t/\hbar} e^{-\beta H_{bA}}\right\} \comma \label{eq:iaw}
\een
where {\boldmath $\mu_A$} is the transition dipole of the acceptor and  
${\bf \hat e} $ is the polarization vector of the radiation.
Inserting the inverse transform of this into Eq. (\ref{eq:kt2}), 
\ben
&&k(t)=\frac{J^2}{\pi\hbar^2 |\mbox{\boldmath $\mu_A$}\cdot {\bf \hat e} |^2}\int_{-\infty}^{\infty} d\omega\ I_A(\omega)\nonumber \\
&&\mbox{\makebox[.2 in]{ }}\times {\rm Re}\left [\int_0^t dt' e^{-i\omega t'+i\epsilon_{_D}t'/\hbar}\frac{1}{Z_{bD}}Tr_{bD}\left\{e^{i(H_{bD}+\bthd)t/\hbar}\right . \right . \nonumber \\
&&\mbox{\makebox[.3 in]{ }}\times\left. \left .e^{-iH_{bD}t'/\hbar}e^{-i(H_{bD}+\bthd)(t-t')/\hbar} e^{-\beta H_{bD}}\right\} \right ]\perd \label{eq:kt3}
\een
The time dependent part involving the donor can be expressed as the 
stimulated emission profile in a pump-probe experiment.  In Appendix
A, we derive a time dependent stimulated emission profile of the donor
which is subject to a stationary field after being excited by a delta pulse.
Inserting Eq. (\ref{eq:edt}) into Eq. (\ref{eq:kt3}), the frequency domain 
expression of the EET rate is given by 
\be
k(t)=\frac{J^2}{2 \pi\hbar^2 |\mbox{\boldmath $\mu_A$}\cdot {\bf \hat e} |^2 |\mbox{\boldmath $\mu_D$}\cdot {\bf \hat e}|^2 }\int_{-\infty}^{\infty}d\omega I_A (\omega) E_D(t,\omega) \perd \label{eq:kt-spov}
\ee
In the limit where $t\rightarrow \infty$, this expression becomes 
equivalent to the F\"{o}rster's  spectral overlap expression  
as long as $E_D(\infty,\omega)$ is equal to the spontaneous emission 
profile of the excited donor except for a normalization factor and the universal frequency 
dependent scaling function.

Equation (\ref{eq:kt-spov}) is the central result of the present 
paper.  It is the pump-probe version of the hot transfer rate\cite{tekhver-jetp42,ht} and 
generalizes the spectral overlap expression of F\"{o}rster
for fast EET processes.   With the modern development of ultrafast 
spectroscopy, determination of $k(t)$ and $E_D(t,\omega)$ can be made 
even in femtosecond scale.  With the advances in experimental 
techniques of altering chromophores by chemical or biological manipulations, 
independent determinations of $I_A(\omega)$ and $E_D(t,\omega)$ can be 
done at the same condition as that of $k(t)$ for a broad range of 
systems.  If these practical issues are settled and if the system 
satisfies the requirements for applying the perturbation theory, 
Eq. (\ref{eq:kt-spov}) should hold as long as the effect of common
modes is insignificant.  

In Eq. (\ref{eq:kt1}), we have defined the reaction rate as 
the time derivative of $P_A(t)$.  Due to the use of perturbation 
theory, such a definition gives a valid result only for $\int_0^t dt' k(t') << 1$.  
In the longer time limit when the population transfer has occurred 
significantly, instead, the reaction rate should be understood as 
the exponential decay rate of $1-P_A(t)$.   The justification for this 
comes from 
the Redfield-type equation\cite{rackovsky-mp25} and the assumption that 
the bath relaxation has been completed.  A reasonable way of combining 
these two limits is to exponentiate the time integration of $k(t)$ as follows:
\be
P_A(t)\approx 1-\exp \left(-\int_0^t dt'\ k(t')\right)  \comma \label{eq:pat}
\ee
where it has been assumed that the transfer from $D$ to $A$ is
irreversible due to either energetic or entropic reason.  
Equation (\ref{eq:pat}) is not based on a rigorous 
derivation, and conditions when such approximation is valid need
to be clarified based on a more rigorous approach, which will
be done elsewhere.  However, for the purpose of understanding the qualitative
aspect of EET kinetics in the nonequilibrium situation, which is the main 
purpose of the present paper, the expression of Eq. (\ref{eq:pat}) is useful.

\subsection{Linearly coupled harmonic oscillator bath}
For the simple case where  the bath consists of independent 
harmonic oscillators, an explicit expression can be found 
for $k(t)$.   Assume that the bath Hamiltonian is given 
by
\be
H_b=\sum_n \hbar\omn \left(\ocb+\frac{1}{2}\right) \comma \label{eq:hb-har}
\ee
and the chromophore-bath interaction is linear in the bath 
coordinate as follows:  
\be
H_{eb}=\sum_{n}\hbar\omn (\bn+\bdn)(\gnd\ocd+\gna\oca) \comma \label{eq:heb-har}
\ee
where $\gnd\gna=0$ and thus the condition of Eq. (\ref{eq:nocom2}) is 
satisfied.  For an explicit calculation, it is convenient to 
start from Eq. (\ref{eq:kt2}).  The integrand of the reaction 
rate involves trace of the product of the propagators for displaced 
harmonic oscillators.  
Each trace over the acceptor bath and the donor bath can be 
done explicitly, and the resulting expression for the reaction 
rate can be written as 
\ben
k(t)&=&\frac{2J^2}{\hbar^2}{\rm Re}\left [ \int_0^t dt'\ e^{i(\tden-\taen)t'/\hbar-i\sum_n (\gnd^2+\gna^2) \sin (\omega_n t') } \right .\nonumber \\
&&\mbox{\makebox[.05 in]{ }}\times e^{2i\sum_n \gnd^2(\sin(\omega_n t)- \sin(\omega_n(t-t'))}\nonumber \\
&&\mbox{\makebox[.05 in]{ }}\left . \times e^{-2\sum_n(\gnd^2+\gna^2)\coth(\beta\hbar\omega_n/2)\sin^2(\omega_n t'/2)}\right ] \comma \label{eq:kt4}
\een
where
\ben
\tilde \epsilon_{D(A)}&=&\epsilon_{D(A)}-\sum_n\hbar \omn g_{nD(A)}^2\nonumber \\
&\equiv& \epsilon_{D(A)}-\lambda_{D(A)}\comma
\een
where $\lambda_D$ and $\lambda_A$ are reorganization energies of 
the donor and acceptor baths.
For the harmonic oscillator bath model, in fact, the  reaction rate 
can be calculated explicitly even for more general situation where 
there are common modes coupling both the donor and acceptor.  In 
Appendix B, we derive the general expression employing the 
small polaron transformation.  
Equation (\ref{eq:kt4}) can alternatively be obtained
from Eq. (\ref{eq:b-kt3}) by letting $g_{nD}g_{nA}=0$.

Define the following spectral densities: 
\ben
\eta_D(\omega)&\equiv&\sum_{n} g_{nD}^2\omega_n^2\delta(\omega-\omega_n)\comma \label{eq:etad}\\
\eta_A(\omega)&\equiv&\sum_{n} g_{nA}^2\omega_n^2\delta(\omega-\omega_n)\perd \label{eq:etaa}
\een
Inserting these definitions into Eq. (\ref{eq:kt4}), the nonequilibrium 
EET rate can be expressed as 
\ben
k(t)&&=\frac{2J^2}{\hbar^2}{\rm Re}\left [\int_0^t dt'\ e^{i(\tden-\taen)t'/\hbar}\right .\nonumber \\
&&\times e^{-i\int_0^\infty d\omega (\eta_D(\omega)+\eta_A(\omega))\sin (\omega t')/\omega^2} \nonumber \\
&&\times e^{2i\int_0^\infty d\omega \eta_D(\omega)(\sin(\omega t)-\sin(\omega(t-t')))/\omega^2}\nonumber \\
&&\left . \times e^{-2\int_0^\infty d\omega (\eta_D(\omega)+\eta_A(\omega))\coth(\beta\hbar\omega/2)\sin^2(\omega t'/2)/\omega^2}\right ]\ . 
\nonumber \\ 
\label{eq:kt5}
\een
This is the main result of the present subsection.  
The expression for $k_r(t)$, which can be calculated from 
Eq. (\ref{eq:krt-1}) through a similar procedure is given by 
\ben
k_r(t)&=&\frac{2J^2}{\hbar^2}{\rm Re}\left [\int_0^t dt'\ e^{i(\tden-\taen)t'/\hbar} \right .\nonumber \\
&&\times e^{-i\int_0^\infty d\omega (\eta_D(\omega)+\eta_A(\omega))\sin (\omega t')/\omega^2}\nonumber \\
&&\left . \times e^{-2\int_0^\infty d\omega (\eta_D(\omega)+\eta_A(\omega))\coth(\beta\omega/2)\sin^2(\omega t'/2)/\omega^2}\right ] \ . 
\nonumber \\ 
\label{eq:krt-2}
\een 
Due to the nonstationary initial condition, the integrand in Eq. (\ref{eq:kt5}) has an additional dependence on $t$, which becomes clear when we compare 
the expression with that of $k_r(t)$.
In general, further simplifications of the rate expressions given by
Eqs. (\ref{eq:kt5}) and (\ref{eq:krt-2}) cannot be made.  However, 
this does not pose a practical difficulty in calculating the reaction
rate because direct numerical integrations of these expressions can be
done quite easily.  In the next section, we implement these calculations
for a simple model spectral density.

Before concluding this section, we examine one important limit 
where the stationary phase approximation is possible.  In the strong
excitation-bath coupling or high temperature limit, the
dominant contribution of the integration comes from small $t'$ region.
Expanding all the functions of $t'$ in the exponent up to the second
order, Eq. (\ref{eq:kt5}) can be approximated as
\ben
k_s(t)&=& \frac{2J^2}{\hbar^2}{\rm Re} \left [ \int_0^t dt' e^{i(\delte-\lambda_T+2C(t))t'/\hbar-(D(\beta)/2-iS(t))t'^2}\right ] \nonumber \\
&=&2J^2 {\rm Re} \left [ e^{-(\delte-\lambda_T+2C(t))^2/(2\hbar^2(D(\beta)-2iS(t)))}\right .\nonumber \\
&&\left .\times \int_0^t dt' e^{-(D(\beta)/2-iS(t))\left (t'-i\frac{(\delte -\lambda_T+2C(t))}{\hbar (D(\beta)-2iS(t))}\right )^2}\right ] \label{eq:kst}\comma
\nonumber \\ 
\een
where  
\ben
\lambda_{T}&\equiv& \hbar \int_0^\infty d\omega \frac{\eta_D(\omega)+\eta_A(\omega)}{\omega}=\lambda_D+\lambda_A\comma \\
S(t)&\equiv &\int_0^\infty d\omega \eta_D(\omega)\sin (\omega t)\comma \\
C(t)&\equiv & \hbar \int_0^\infty d\omega \frac{\eta_D(\omega)}{\omega}\cos(\omega t)\comma\\
D(\beta)&\equiv&\int_0^\infty d\omega (\eta_D(\omega)+\eta_A(\omega))\coth(\frac{\beta \hbar \omega}{2}) \perd
\een
Although Eq. (\ref{eq:kst}) involves a simpler integration 
than Eq. (\ref{eq:kt5}), the complex Gaussian integration 
for finite $t$ does not convey a clear physical picture. 
If $|S(t)| << D(\beta)$ and in the long enough time limit, 
one can disregard the imaginary term $S(t)$ and approximate 
the integration in Eq. (\ref{eq:kst}) as that from $0$ to $\infty$.  
The rate expression under this situation can be simplified as 
\be
k_{sg}(t)= \frac{J^2}{\hbar^2}\sqrt{\frac{2\pi}{D(\beta)}} e^{-(\delte-\lambda_T+2C(t))^2/2\hbar^2 D(\beta)} \perd \label{eq:ksgt}
\ee
This approximation is valid only for long time
regime and for strong enough excitation-bath coupling.  The 
maximum rate based on this expression is obtained for 
$\delte=\lambda_T-2C(t)$.  Since $C(\infty)=0$ for most realistic 
spectral densities (cf. sub-Ohmic, $C(\infty)\neq 0$), this relation 
implies that the optimum energy difference in the long time limit is 
given by $\delte=\lambda_T$.  However, the above expression
suggests that the reaction rate for a nonoptimum $\delte$ can also 
be substantial during the transient period 
when $C(t)$ changes.  If $J$ is not small compared to the 
bath relaxation rate, the contribution of this transient term 
cannot be neglected.  A similar observation was made in the 
theory of nonequilibrium electron transfer reaction\cite{cho-jcp103}.

\section{Model calculations}
Assume that the bath modes coupled to $D$ and $A$  have the same 
spectral profile, while their magnitudes of coupling can 
be different.  We consider a model where the number of modes 
increases linearly for small $\omega$ and
has an exponential tail for large $\omega$.  The corresponding 
spectral densities, defined by Eqs. (\ref{eq:etad}) and (\ref{eq:etaa}),
thus have the following form:
\ben
\eta_{_{D(A)}}(\omega)&=&\alpha_{_{D(A)}}\frac{\omega^3}{\omega_c^2}e^{-\omega/\omega_c} \comma  \label{eq:spd}
\een
where $\alpha_{_D}$ and $\alpha_{_A}$ determine the coupling strengths
of $|D\rangle $ and $|A\rangle$ to their respective bath modes.  $\omega_c$ 
is the cutoff frequency which determines the spectral range of the bath.
Due to the common value of this cutoff frequency, the bath relaxation 
dynamics has a single time scale and the reaction rate exhibits a simple 
scaling behavior.  In Appendix C, we provide a detailed form of the reaction 
rate in the units where $\hbar=1$.  As can be seen from Eq. (\ref{eq:kt6}),
all the EET kinetics for different $\omega_c$ can be deduced from the 
following dimensionless rate:
\be
\kappa (\tau)=\frac{\omega_c}{J^2}k(t)\comma\ \tau=\omega_c t \label{eq:kappa}
\ee
For the case where the reaction rate is time dependent, the 
instantaneous value of the reaction rate cannot be a clear 
indication of the efficiency of the energy transfer.  For this
reason, we also provide the following dimensionless quantity:
\be
p(\tau)=\int_0^\tau d\tau' \kappa (\tau') \perd \label{eq:pop}
\ee
According to Eq. (\ref{eq:pat}), $P_A(t)\approx 1-\exp\{-(J/\omega_c)^2 p(\tau)\}$, 
the population of $|A\rangle$ at time $t=\tau/\omega_c$. 

For the reaction rate $k_r(t)$ given by Eq. (\ref{eq:krt-2}),
similar quantities can be defined and these are denoted as $\kappa_r(\tau)$ 
and $p_r (\tau)$.  The expression for $\kappa_r(\tau)$ can be obtained from
that of $\kappa (\tau)$ simply disregarding the third term 
in Eq. (\ref{eq:phi}).  Equations (\ref{eq:lambda-a})-(\ref{eq:dbeta-a}) in 
combination with Eqs. (\ref{eq:kst}) and (\ref{eq:ksgt}) provide explicit 
expressions for $k_s(t)$ and $k_{sg}(t)$.  The dimensionless versions of 
these rates are denoted as $\kappa_s(\tau)$ and $\kappa_{sg}(\tau)$, and 
their cumulative integrations are represented by $p_s(\tau)$ and 
$p_{sg}(\tau)$. 
 
\subsection{Zero temperature limit}
We first performed calculations for $\alpha_D=\alpha_A=10$, a strong coupling 
limit.  The values of the renormalized energy difference $\delte$ 
were chosen to be $\lambda_T-\lambda_D$, 
$\lambda_T$, and $\lambda_T+\lambda_D$, which were motivated by the
form of Eq. (\ref{eq:ksgt}).  These three cases are analogues of 
the normal, activationless, and inverted regimes in the electron transfer
kinetics.  Figure 1 shows the rates.
The nonequilibrium effect is shown to be important until about 
$\tau\approx 2$.  This is in contrast to the behavior of $\kappa_r (\tau)$,
which approaches the long time limit in about $\tau\approx 0.2$.
During the nonstationary period, $\kappa (\tau)$ goes through a 
maximum before approaching the long time limit in a smooth fashion.
The maximum value appears earlier for smaller value of $\delte$, for 
which it takes shorter for the bath to relax into the resonance condition. 
In the classical limit, this can be understood in terms of the average 
time for the wavepacket in the excited state to go through the 
crossing point.

The two approximations of $\kappa_s(\tau)$ and $\kappa_{sg}(\tau)$
reproduce the long time limit quite well for the present case of 
strong coupling limit.  In the nonstationary regime, they exhibit 
a qualitative behavior similar to that of $\kappa(\tau)$, but the
quantitative agreement is not satisfactory in the region of $1 < \tau < 2$.
This indicates that, in this intermediate time regime, the coherence
of the bath still plays a substantial role and cannot be well 
approximated by the stationary phase integration.

In Fig. 2, $p(\tau)$ given by Eq. (\ref{eq:pop}) and analogous 
quantities for other approximations are provided.  
For $\delte=\lambda_T-\lambda_D$, the nonequilibrium contribution
results in larger population transfer during the short initial
period, but not in the longer time period.  For $\delte=\lambda_T$, 
the nonequilibrium effect always leads to smaller population 
transfer.  For $\delte=\lambda_T+\lambda_D$, the nonequilibrium
population transfer is smaller than the fully relaxed one 
until about $\tau\approx 1$, but it becomes larger during the longer 
time as the delayed response of the bath brings a resonance condition.  
The different kinetics during the nonstationary period
is seen to result in finite differences in the net population 
transfer in the long time regime.  The values of $p_s(\tau)$ 
and $p_{sg}(\tau)$ are seen 
to overestimate $p(\tau)$ when $\delte=\lambda_T-\lambda_D$ and 
underestimates it for $\delte=\lambda_T+\lambda_D$.  The stationary 
phase approximation seems to work relatively well for the 
case of $\delte=\lambda_T$.

Some of the patterns observed from the above calculations may be 
specific for the model of the bath and the temperature.  
However, the features that the nonequilibrium effect is substantial until 
about up to $\tau\approx 2$ and that the maximum peak of the reaction rate
appears earlier for smaller $\delte$ are expected to hold quite generally.  
Due to the assumption of strong coupling, the approximate expressions of 
$\kappa_s(\tau)$ and $\kappa_{sg}(\tau)$ agree quite well with 
$\kappa (\tau)$, although there is still a difference in the net population.
Such agreement cannot be seen for the weak coupling case considered next, an
expected result considering the assumptions involved in $\kappa_s (\tau)$ 
and $\kappa_{sg}(\tau)$.

Calculations were performed for $\alpha_D=1$ and $\alpha_A=1$, 
a weak coupling limit.  Figure 3 shows the dimensionless reaction 
rates and Fig. 4 their cumulative integrations.  Unlike the case
of strong coupling, the difference between $\kappa (\tau)$ and 
$\kappa_r(\tau)$ is not appreciable except for some phase difference
in their oscillations.  The resulting time integrations, $p(\tau)$
and $p_r(\tau)$, are quite close as can be seen from Fig. 4.   Especially
for the two cases of $\delte=\lambda_T$ and $\delte=\lambda_T+\lambda_D$, 
the two curves almost overlap with each other.  The approximations 
of  $\kappa_s(\tau)$ and $\kappa_{sg}(\tau)$ are not expected to work 
well for the present weak coupling regime, but it is meaningful to 
examine their qualitative nature.  In Fig. 3 , it is shown that 
$\kappa_s (\tau)$ and $\kappa_{sg}(\tau)$ do not reproduce the oscillatory 
behavior and deviate from $\kappa (\tau)$ systematically, except for the 
case of $\delte=\lambda_T$.  The resulting values of $p_s(\tau)$ 
and $p_{sg}(\tau)$ underestimate $p(\tau)$ for $\delte=\lambda_T-\lambda_D$ 
and overestimate it for $\delte=\lambda_T+\lambda_D$.  These are opposite 
to the trends observed for strong coupling limit.  In conclusion, for the 
present case of weak coupling, the nonequilibrium effect is unimportant, 
and both $\kappa (\tau)$ and $\kappa_r(\tau)$ rise to their steady 
state limit in a similar fashion.  The oscillations persistent in both $\kappa (\tau)$ 
and $\kappa_r(\tau)$ in Fig. 3 originate from $e^{i(\epsilon_{_D}-\epsilon_{_A})t'/\hbar}$ 
in Eqs. (\ref{eq:kt5}) and (\ref{eq:krt-2}), which remains significant even in the 
long time limit due to the weak excitation-bath coupling.  However, even though 
all the assumptions of the present section hold, actual observation of these 
oscillations in a real system is unlikely due to the energy uncertainty related 
dephasing.

\subsection{Temperature effect}
As the temperature increases, the phonon side band becomes broader 
in the spectral overlap expression of Eq. (\ref{eq:kt-spov}).  
The relaxation and dephasing of the bath become faster also.  
It is interesting to examine how these changes affect the nonequilibrium
kinetics.  For this purpose, we compare only $\kappa (\tau)$ and 
$\kappa_r (\tau)$.  The temperature dependence was accounted for 
by using an approximate expression for $\gamma (\tau)$, in Eq. (\ref{eq:kt6}), as follows:
\ben
\gamma(\tau)&\approx& (\alpha_D+\alpha_A)\left (\frac{\tau^2}{1+\tau^2}+\frac{2\tau^2}{(1+\tau^2)^2} \right . \nonumber \\
 &&\left . + \frac{16\tau^2}{\beta\omega_c (\beta\omega_c+2)((\beta\omega_c+2)^2+4\tau^2)} \right )\perd
\een
This is based on the approximation of $\coth (x/2)\approx 1+(2/x)e^{-x/2}$, which reproduces the proper low and high temperature limits.    
Although the approximation is rather crude  in the region of $\beta\omega_c \sim 1$, the 
overall performance is good enough to produce a correct qualitative trend.

In Fig. 5, we provide results for $\alpha_D=\alpha_A=10$, the strong 
coupling limit, and for $\delte=\lambda_T-\lambda_D$.  The zero temperature 
result is shown in the top panel of Fig. 1.  As the temperature 
increases, the nonequilibrium effects diminish.  However, even for very 
high temperature limit of $\beta\omega_c=0.1$, there is a noticeable difference
between $\kappa (\tau)$ and $\kappa_r(\tau)$.  Only in the highest 
temperature limit of $\beta\omega_c=0.01$, $\kappa (\tau)$ and 
$\kappa_r(\tau)$ show an agreement.  This indicates that for 
strongly coupled systems, the nonequilibrium effects persist even 
up to very high temperature.  
Next, we considered the weak coupling case of $\alpha_D=\alpha_A=1$ 
with the same choice of $\delte=\lambda_T-\lambda_D$.  
The results are shown in Fig. 6.
In the zero temperature limit, the nonequilibrium effect was not 
so significant for this weak coupling case.  
This fact remains the same even for the higher temperature.

\section{Discussion}
The main focus of the present paper was the nonequilibrium 
bath relaxation effect.  For this reason, we considered only 
the simplest case where donor and acceptor each have one excited
state.  However, as recent studies on various systems 
indicate\cite{ret,hu-pt,sundstrom-jpcb103,renger-pr343}, the 
existence of multiple chromophores is a general situation rather than 
an exception.  Therefore, one in general must consider 
the case where the donor and the acceptor respectively consist 
of multiple exciton states.
Given that the inter-excitonic relaxation can be disregarded
or approximated in a self consistent way, the generalization 
of the present theory can be done straightforwardly  
as are those of the FD theory\cite{sumi-jpcb103,mukai-jpcb-103,mukai-jl87-89,scholes-jpcb104,scholes-jpcb105,damjanovic-pre59,yeow-jpca104}.  In many cases, this type of simple extension 
may indeed capture the main aspect of the EET between multichromorphic 
donor and acceptor.   However, if some of the excitonic levels
are closely spaced or if there is degeneracy due to symmetry, 
more careful theoretical analysis is necessary.

In a heterogeneous environment or a complex system, disorder
plays an important role and its explicit consideration becomes
necessary for a proper understanding of the system.  Due to the 
averaging over the static disorder involved in this case, 
our spectral overlap expression cannot be used directly for the 
observables of an ensemble experiment.   However, given that 
the distribution of the static disorder is well characterized 
by other means, Monte Carlo simulation can provide an indirect 
way of determining the nonequilibrium population transfer.  
The study of the interplay between the heterogeneity and the 
nonequilibrium effect in this way can bring new insights into the 
fast EET kinetics of complex systems.  

In the LH2 of purple bacteria, the EET from B800 to B850 occurs 
in about $1 {\rm ps}$ and the rate is not so sensitive to temperature\cite{sundstrom-jpcb103}.  
This rate is much faster than that obtained by the FD theory, but it 
has recently been shown that consideration of the multichromorphic 
nature of the B850 and the disorder account for much of the discrepancy\cite{sumi-jpcb103,mukai-jpcb-103,mukai-jl87-89,scholes-jpcb104}. 
However, the theoretical value is still somewhat smaller than the experimental 
one\cite{scholes-jpcb104}.  Many explanations are possible for this, and the nonequilibrium effect is 
one such possibility. 

According to our model calculations, the nonequilibrium short time
kinetics shows complicated time dependent behavior and the reaction
rate in this regime is less dependent on the spectral overlap between 
the stationary emission and absorption profiles.  Some of these features
can be seen in recent sub-picosecond pump-probe experiments\cite{cerullo-cpl335,neuwahl-jpcb105,king-jpcb105}.  
For example, the reaction rate is shown to be relatively insensitive
to the change of the spectral profile for the EET in the photosynthetic
reaction center\cite{king-jpcb105}.

Before applying the present theory to a given system, 
it is always important to examine 
whether the nature of the system is consistent with the assumption 
of irreversibility and the use of second order approximation.  
It is expected that our results can be applied as long as $\delte$ is 
sufficiently large and the bath relaxation of the acceptor is 
fast enough.  However, in order to address these issues more 
systematically, it is necessary to formulate a theory that can account 
for electronic coherence.   For this purpose, one may adopt the formalism 
of the generalized master equation\cite{zhang-jcp108} 
or consider in the framework of the Redfield-type 
equation\cite{rackovsky-mp25}. 

\acknowledgements 
SJ thanks Prof. R. M. Hochstrasser for discussions on the EET and pump-probe 
experiments.  

\appendix
\section{Time dependent stimulated emission profile of the donor}
Consider a system consisting of the donor and its own bath.  
The relevant Hamiltonian in the single excitation manifold is
given by
\be
H_{D}=\den a_{_D}^{\dagger}a_{_D}+B_{_D}a_{_D}^{\dagger}a_{_D}+H_{b_D} \perd
\ee
Assume that the donor is excited at time zero by a delta pulse.  
Then, the initial density operator at $t=0$ is given by
\be
\rho (0)=|D\rangle \langle D|\frac{1}{Z_{b_D}}e^{-\beta H_{b_D}} \perd
\ee
Right after the pulse excitation, a stationary field with a fixed
frequency $\omega$ is turned on.  The time dependent Hamiltonian governing 
the dynamics for $t > 0$, assuming unit field strength, is given by 
\be
H(t)=H_D+|\mbox{\boldmath $\mu_D$}\cdot {\bf \hat e}|(e^{-i\omega t}|D\rangle \langle g|+e^{i\omega t}|g\rangle \langle D|) \comma
\ee
where rotating wave approximation was used.
The probability for the donor to emit a photon 
and go to the ground state, induced by the stationary field,
is given by 
\ben
&&P_g(t)=|\mbox{\boldmath $\mu_D$}\cdot {\bf \hat e}|^2\int_0^t dt'\int_0^t dt''\ e^{i\omega(t'-t'')-i\den(t'-t'')/\hbar} \nonumber \\
&&\mbox{\makebox[.5 in]{ }}\times \frac{1}{Z_{b_D}} Tr_{b_D} \left\{ e^{-iH_{b_D}(t-t')/\hbar }e^{-i(B_D+H_{b_D})t'/\hbar} \right .\nonumber \\
&&\mbox{\makebox[.6 in]{ }}\left . \times\ e^{-\beta H_{b_D}}e^{i(B_D+H_{b_D})t''/\hbar} e^{iH_{b_D}(t-t'')/\hbar}\right\} \comma
\een
where weak field approximation was used.
The time dependent stimulated emission profile is defined as the time 
derivative of this probability as follows:
\ben
E_D(t,\omega)&\equiv& \frac{d}{dt}P_g(t) \nonumber \\
&=&2|\mbox{\boldmath $\mu_D$}\cdot {\bf \hat e}|^2 {\rm Re}\left [\int_0^t dt' e^{-i\omega t'+i\den t'/\hbar} \right . \nonumber \\
&&\times \frac{1}{Z_{b_D}}Tr_{b_D}\left\{e^{i(H_{b_D}+B_D)t/\hbar}e^{-iH_{b_D}t'/\hbar}\right . \nonumber \\
&&\mbox{\makebox[.2 in]{ }}\left . \left . \times\ e^{-i(H_{b_D}+B_D)(t-t')/\hbar} e^{-\beta H_{b_D}}\right\}\right] \perd \label{eq:edt}
\een 
   
\section{rate expression for a general harmonic oscillator bath}
For a general harmonic oscillator bath model, we derive the 
expression of the nonequilibrium reaction rate given by 
Eq. (\ref{eq:kt1}) based on the small polaron transformation.
For this purpose, first we introduce the following generator 
of small polaron transformation:
\be
S=-\sum_n (\bn-\bdn)(\gnd \ocd+\gna\oca ) \perd
\ee 
Then, it is straightforward to show that
\ben
\tilde{H}=e^S H e^{-S}&=&\tden\ocd +\taen \oca +H_b+\tilde{H}_{DA}\nonumber \\
&=&\tilde{H}_0 + \tilde{H}_{DA} \comma
\een
where
\ben
{\tilde H}_{DA}&=&Je^S (\odd\oa+\oda\od )e^{-S}\nonumber \\
            &=&J(\thdd \tha \odd\oa +\thda \thd \oda\od)\comma
\een
where
\be
\theta_{D(A)}^\dagger=e^{-\sum_n g_{nD(A)}(\bn-\bdn)} \perd
\ee
Inserting $1=e^{-S}e^S$ between every two operators in Eq. (\ref{eq:kt1}),
one can show that
\ben
k(t)&=&\frac{2J^2}{\hbar^2}{\rm Re}\left [\int_0^t dt' e^{i(\tden-\taen)t'/\hbar}\right .\nonumber \\
&&\left . \times\frac{1}{Z_b} Tr_b\left\{e^{iH_bt'/\hbar}\Theta^\dagger e^{-iH_bt'/\hbar}\Theta \rho_d(t-t')\right\}\right ]\comma \label{eq:a-kt-1}
\een 
where $\Theta^\dagger=\thdd\tha$,  
\ben
\rho_d(t)&\equiv& e^{-iH_bt/\hbar}\thdd e^{-\beta H_b}\thd e^{iH_bt/\hbar}\nonumber \\
&=&e^{-\beta\sum_n \hbar\omega_n(\tilde b_{nD}^\dagger (t)\tilde b_{nD}(t) +\frac{1}{2})}\comma\label{eq:a-rhodt}
\een
with
\be
\tilde b_{nD}(t)\equiv b_n-g_{nD}e^{-i\omega_n t} \perd \label{eq:a-bnd}
\ee
In Eq. (\ref{eq:a-kt-1}),
\ben
&&e^{iH_bt/\hbar}\Theta^\dagger e^{-iH_bt/\hbar}\nonumber \\
&&=\exp\left\{-\sum_n g_{n\Delta }(\bn e^{-i\omega_n t}-\bdn e^{i\omega_n t})\right\}\comma  \label{eq:a-t-theta}
\een
where $g_{n\Delta}\equiv g_{nD}-g_{nA}$.
Inserting Eqs. (\ref{eq:a-rhodt})-(\ref{eq:a-t-theta}) into Eq. (\ref{eq:a-kt-1}),
\ben
k(t)&=&\frac{2J^2}{\hbar^2}{\rm Re}\left [\int_0^t dt' e^{i(\tden-\taen)t'/\hbar} e^{-i\sum_n \gndt \sin (\omega_n t')}\right .\nonumber \\
&&\mbox{\makebox[.1 in]{ }}\times e^{2i\sum_n \gndt \gnd(\sin(\omega_n t)-\sin(\omega_n (t-t')))} \nonumber \\
&&\mbox{\makebox[.1 in]{ }}\times \frac{1}{Z_b}Tr_b\left\{e^{-\sum_n \gndt ((e^{-i\omega_nt'}-1)\tilde b_{nD} -(e^{i\omega_nt'}-1)\tilde b_{nD}^\dagger) }\right . \nonumber \\ 
&&\mbox{\makebox[.3 in]{ }}\left . \times\left . e^{-\beta \sum_n \hbar\omega_n (\tilde b_{nD}^\dagger \tilde b_{nD}+\frac{1}{2})}\right \}\right ] \comma \label{eq:a-kt-2}
\een 
where the time dependences of $\tilde b_{nD}(t-t')$ and 
$\tilde b_{nD}^\dagger(t-t')$ have been omitted, but this does not 
make any difference in the result because of the trace operation.  
Disentangling the summation in the exponent and then evaluating 
the trace over the bath, one can reduce Eq. (\ref{eq:a-kt-2}) into 
the following integral expression: 
\ben
k(t)&=&\frac{2J^2}{\hbar^2}{\rm Re}\left [ e^{2i\sum_n \gndt \gnd \sin(\omega_n t)}\int_0^t dt'\ e^{i(\tden-\taen)t'/\hbar} \right .\nonumber \\
&&\mbox{\makebox[.2 in]{ }}\times e^{-i\sum_n \gndt^2 \sin (\omega_n t')-2i\sum_n \gndt\gnd \sin(\omega_n(t-t'))}\nonumber \\
&&\mbox{\makebox[.2 in]{ }}\left . \times e^{-2\sum_n\gndt^2\coth(\beta\hbar\omega_n/2)\sin^2(\omega_n t'/2)}\right ] \perd \label{eq:b-kt3}
\een

For the case where there is no common mode, $g_{n\Delta}g_{nD}=g_{nD}^2$
and $g_{n\Delta}^2=g_{nD}^2+g_{nA}^2$ and the expression of Eq. (\ref{eq:kt4})
is reproduced.

\section{Rate expressions for the model spectral densities of Sec. III}
For the spectral densities given by Eq. (\ref{eq:spd}), the integrations
within the exponents of Eq. (\ref{eq:kt5}) can be performed explicitly.
Adopting the units where $\hbar=1$, the reaction rate of Eq. (\ref{eq:kt5})
can be expressed as
\be
k(t)=\frac{2J^2}{\omega_c} {\rm Re}\left [\int_0^{\tau} d\tau'\ e^{i\phi(\tau,\tau')-\gamma (\tau')}\right]\comma \label{eq:kt6}
\ee 
where $\tau=\omega_c t, \tau'=\omega_c t'$, and
\ben
\phi(\tau,\tau')&=& \left (\frac{\den-\aen}{\omega_c}-2 (\alpha_{_D}-\alpha_{_A})\right)\tau'-\frac{2(\alpha_{_D}+\alpha_{_A})\tau'}{(1+\tau'^2)^2}\nonumber \\
&&-4\alpha_{_D}\left(\frac{\tau-\tau'}{(1+(\tau-\tau')^2)^2}-\frac{\tau}{(1+\tau^2)^2}\right) \perd \label{eq:phi}
\een
In Eq. (\ref{eq:kt6}), the function $\gamma(\tau')$ comes from 
the integration involving $\coth (\beta\omega \hbar/2)$ in Eq. (\ref{eq:kt5}).
Simple approximation for this function is possible in the limits of low and 
high temperature.  Using these approximations, 
\ben
\gamma(\tau)&&=\left\{ \begin{array}{c}
              (\alpha_{_D}+\alpha_{_A})\left(\frac{\tau^2}{1+\tau^2}+\frac{2\tau^2}{(1+\tau^2)^2}\right) \ ,\ \beta\omega_c >> 1 \comma \\    
             \frac{2(\alpha_{_D}+\alpha_{_A})}{\omega_c\beta}\frac{\tau^2}{1+\tau^2}\ , \ \beta\omega_c << 1 \perd
\end{array} \right . \label{eq:gamma}
\een
In the main text, we consider the zero temperature limit first and 
study the finite temperature effect using a simple interpolation formula.
Similarly, the quantities entering the approximate rate expressions of
Eqs. (\ref{eq:kst}) and (\ref{eq:ksgt}) can be calculated.  These are 
\ben
\lambda_{T}&=&2(\alpha_{_D}+\alpha_{_A}) \omega_c \comma\\
S(t)&=&24 \alpha_{_D}\omega_c^2\frac{\tau-\tau^3}{(1+\tau^2)^4} \comma\ \tau=\omega_c t\comma \label{eq:lambda-a}\\
C(t)&=&2\alpha_{_D}\omega_c \frac{1-\tau^2}{(1+\tau^2)^3} \comma\ \tau=\omega_c t \comma \\
D(\beta)&=&\left\{ \begin{array}{c}
              6(\alpha_{_D}+\alpha_{_A})\omega_c^2\ ,\ \beta\omega_c >> 1 \comma \\                         4(\alpha_{_D}+\alpha_{_A})\omega_c/\beta\ , \ \beta\omega_c << 1 \perd
\end{array} \right . \label{eq:dbeta-a}
\een


\begin{thebibliography}{10}

\bibitem{forster-dfs}
{Th. F\"{o}rster},
\newblock Discuss. Faraday Soc. {\bf 27}, 7 (1953).

\bibitem{forster-book}
{Th. F\"{o}rster} , in {\it Modern Quantum Chemistry, Part III }, edited by O.
  Sinanoglu (Academic Press, New York, 1965).

\bibitem{dexter-jcp}
D.~L. Dexter,
\newblock J. Chem. Phys. {\bf 21}, 836 (1953).

\bibitem{agranovich}
V.~M. Agranovich and M.~D. Galanin,
\newblock {\em Electronic excitation energy transfer in condensed matter},
\newblock North-Holland, Amsterdam, 1982.

\bibitem{eet-org}
{P. Reineker, H. Haken, and H. C. Wolf, Ed.},
\newblock {\em Organic Molecular Aggregates},
\newblock Springer-Verlag, Berlin, 1983.

\bibitem{birks}
E.~J.~B.~Birks,
\newblock {\em Excited States of Biological Molecules},
\newblock John Wiley \& Sons, London, 1976.

\bibitem{ret}
D.~L. Andrews and A.~A. Demidov,
\newblock {\em Resonance Energy Transfer},
\newblock John Wiley \& Sons, Chichester, 1999.

\bibitem{cerullo-cpl335}
{G. Cerullo, S. Stagira, M. Zavelani-Rossi, S. De Silvestri, T. Virgili, D. G.
  Lidzey, and D. D. C. Bradley},
\newblock Chem. Phys. Lett. {\bf 335}, 27 (2001).

\bibitem{meskers-cpl339}
{S. C. J. Meskers, J. H\"{u}bner, M. Oestreich, and H. B\"{a}ssler},
\newblock Chem. Phys. Lett. {\bf 339}, 223 (2001).

\bibitem{buckley-cpl339}
{A. R. Buckley, M. D. Rahn, J. Hill, J. Cabanillas-Gonzalez, A. M. Fox, and D.
  D. C. Bradley},
\newblock Chem. Phys. Lett. {\bf 339}, 331 (2001).

\bibitem{neuwahl-jpcb105}
{F. V. R. Neuwahl, R. Righini, A. Adronov, P. R. L. Malenfant, and J. M. J.
  Fr\'{e}chrt},
\newblock J. Phys. Chem. B {\bf 105}, 1307 (2001).

\bibitem{Morgado-jmc11}
{J. Morgado, F. Cacialli, R. Iqbal, S. C. Moratti, A. B. Holmes, G. Yahioglu,
  L. R. Milgrom, and R. H. Friend},
\newblock J. Mater. Chem. {\bf 11}, 278 (2001).

\bibitem{hu-pt}
X.~Hu and K.~Schulten,
\newblock Phys. Today {\bf August}, 28 (1997).

\bibitem{sundstrom-jpcb103}
{V. Sundstrom, T. Pullerits, and R. van Grondelle},
\newblock J. Phys. Chem. B {\bf 103}, 2327 (1999).

\bibitem{renger-pr343}
{T. Renger, V. May, and O. K\"{u}hn},
\newblock Phys. Rep. {\bf 343}, 137 (2001).

\bibitem{soules-prb3}
T.~F. Soules and C.~B. Duke,
\newblock Phys. Rev. B {\bf 3}, 262 (1971).

\bibitem{rackovsky-mp25}
S.~Rackovsky and R.~Silbey,
\newblock Mol. Phys. {\bf 25}, 61 (1973).

\bibitem{jackson-jcp78}
B.~Jackson and R.~Silbey,
\newblock J. Chem. Phys. {\bf 78}, 4193 (1983).

\bibitem{silbey-arpc27}
R.~Silbey,
\newblock Annu. Rev. Phys. Chem. {\bf 27}, 203 (1976).

\bibitem{grover-jcp54}
M.~Grover and R.~Silbey,
\newblock J. Chem. Phys. {\bf 54}, 4843 (1971).

\bibitem{munn-jcp68}
R.~W. Munn and R.~Silbey,
\newblock J. Chem. Phys. {\bf 68}, 2439 (1978).

\bibitem{juzeliunas}
G. Juzeliunas and D. L. Andrews in Ref. 7.

\bibitem{hsu-jcp114}
{C.-P. Hsu, G. R. Fleming, M. Head-Gordon, and T. Head-Gordon},
\newblock J. Chem. Phys. {\bf 114}, 3065 (2001).

\bibitem{sumi-jpcb103}
H.~Sumi,
\newblock J. Phys. Chem. B {\bf 103}, 252 (1999).

\bibitem{mukai-jpcb-103}
K.~Mukai, S.~Abe, and H.~Sumi,
\newblock J. Phys. Chem. B {\bf 103}, 6096 (1999).

\bibitem{mukai-jl87-89}
K.~Mukai, S.~Abe, and H.~Sumi,
\newblock J. Lumin. {\bf 87-89}, 818 (2000).

\bibitem{scholes-jpcb104}
G.~D. Scholes and G.~R. Fleming,
\newblock J. Phys. Chem. B {\bf 104}, 1854 (2000).

\bibitem{scholes-jpcb105}
G.~D. Scholes, X.~J. Jordanides, and G.~R. Fleming,
\newblock J. Phys. Chem. B {\bf 105}, 1640 (2001).

\bibitem{damjanovic-pre59}
{A. Damjanovi\'{c}, T. Ritz, and K. Schulten},
\newblock Phys. Rev. E {\bf 59}, 3293 (1999).

\bibitem{yeow-jpca104}
E.~K.~L. Yeow and K.~P. Ghiggino,
\newblock J. Phys. Chem. B {\bf 104}, 5825 (2000).

\bibitem{lin-pre}
S.~H. Lin, W.~Z. Xiao, and W.~Dietz,
\newblock Phys. Rev. E. {\bf 47}, 3698 (1993).

\bibitem{kakitani-jpcb103}
T.~Kakitani, A.~Kimura, and H.~Sumi,
\newblock J. Phys. Chem. B {\bf 103}, 3720 (1999).

\bibitem{kimura-jl87-9}
A.~Kimura, T.~Kakitani, and T.~Yamato,
\newblock J. Lumin. {\bf 87-89}, 815 (2000).

\bibitem{kimura-jpcb104}
A.~Kimura, T.~Kakitani, and T.~Yamato,
\newblock J. Phys. Chem. B {\bf 104}, 9276 (2000).

\bibitem{tekhver-jetp42}
I.~Y. Tekhver and V.~V. Khizhnyakov,
\newblock Sov. Phys.-JETP {\bf 42}, 305 (1976).

\bibitem{ht}
See pages 149-150 of Ref. 4 for a comprehensive list of references.

\bibitem{sumi-jpsj51}
H.~Sumi,
\newblock J. Phys. Soc. Japan {\bf 51}, 1745 (1982).

\bibitem{sumi-prl50}
H.~Sumi,
\newblock Phys. Rev. Lett. {\bf 50}, 1709 (1983).

\bibitem{gnanakaran}
S. Gnanakaran, G. Haran, R. Kumble, and R. M. Hochstrasser in Ref. 7.

\bibitem{vangrondelle}
R. van Grondelle and O. J. G. Somsen in Ref. 7.

\bibitem{king-jpcb105}
{B. A. King, A. de Winter, T. B. McAnaney, and S. G. Boxer},
\newblock J. Phys. Chem. B {\bf 105}, 1856 (2001).

\bibitem{cho-jcp103}
M.~Cho and R.~J. Silbey,
\newblock J. Chem. Phys. {\bf 103}, 595 (1995).

\bibitem{zhang-jcp108}
{W. M. Zhang, T. Meier, V. Chernyak, and S. Mukamel},
\newblock J. Chem. Phys. {\bf 108}, 7763 (1998).

\end{thebibliography}

\end{multicols}

\begin{figure}
\epsfxsize=6 in
\epsffile{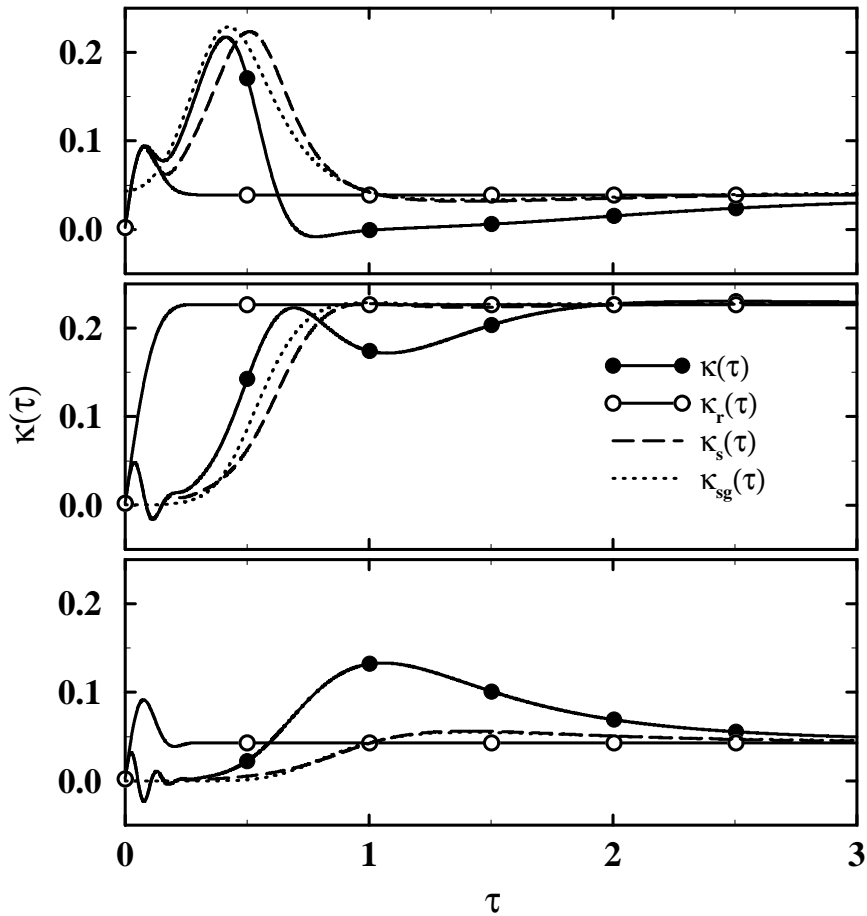}
\caption{Dimensionless nonequilibrium reaction rate $\kappa (\tau)$, Eq. (\ref{eq:kappa}), and analogous quantities for $k_r(t)$, $k_s(t)$, and $k_{sg}(t)$.
$\alpha_{_D}=\alpha_{_A}=10$ and temperature is zero.  The top panel corresponds to $\delte=\lambda_T-\lambda_D$, the middle panel to $\delte=\lambda_T$, and the bottom panel to $\delte=\lambda_T+\lambda_D$.} 
\end{figure}

\begin{figure}
\epsfxsize=6 in
\epsffile{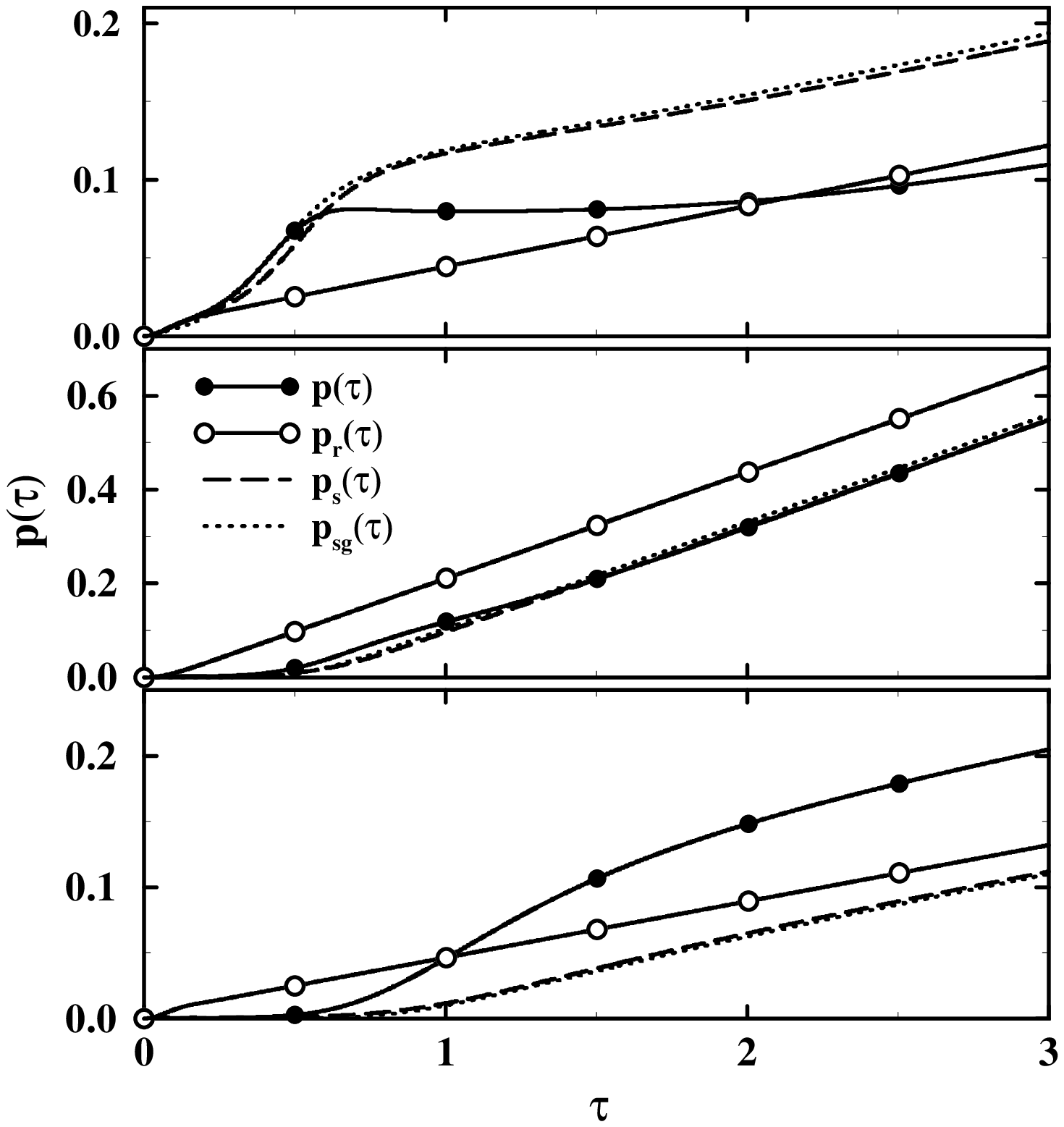}
\caption{Scaled population $p(\tau)$, Eq. (\ref{eq:pop}), and analogous quantities
for  $k_r(t)$, $k_s(t)$, and $k_{sg}(t)$. 
$\alpha_{_D}=\alpha_{_A}=10$ and temperature is zero.  The top panel corresponds to $\delte=\lambda_T-\lambda_D$, the middle panel to $\delte=\lambda_T$, and the bottom panel to $\delte=\lambda_T+\lambda_D$.}
\end{figure}

\begin{figure}
\epsfxsize=6 in
\epsffile{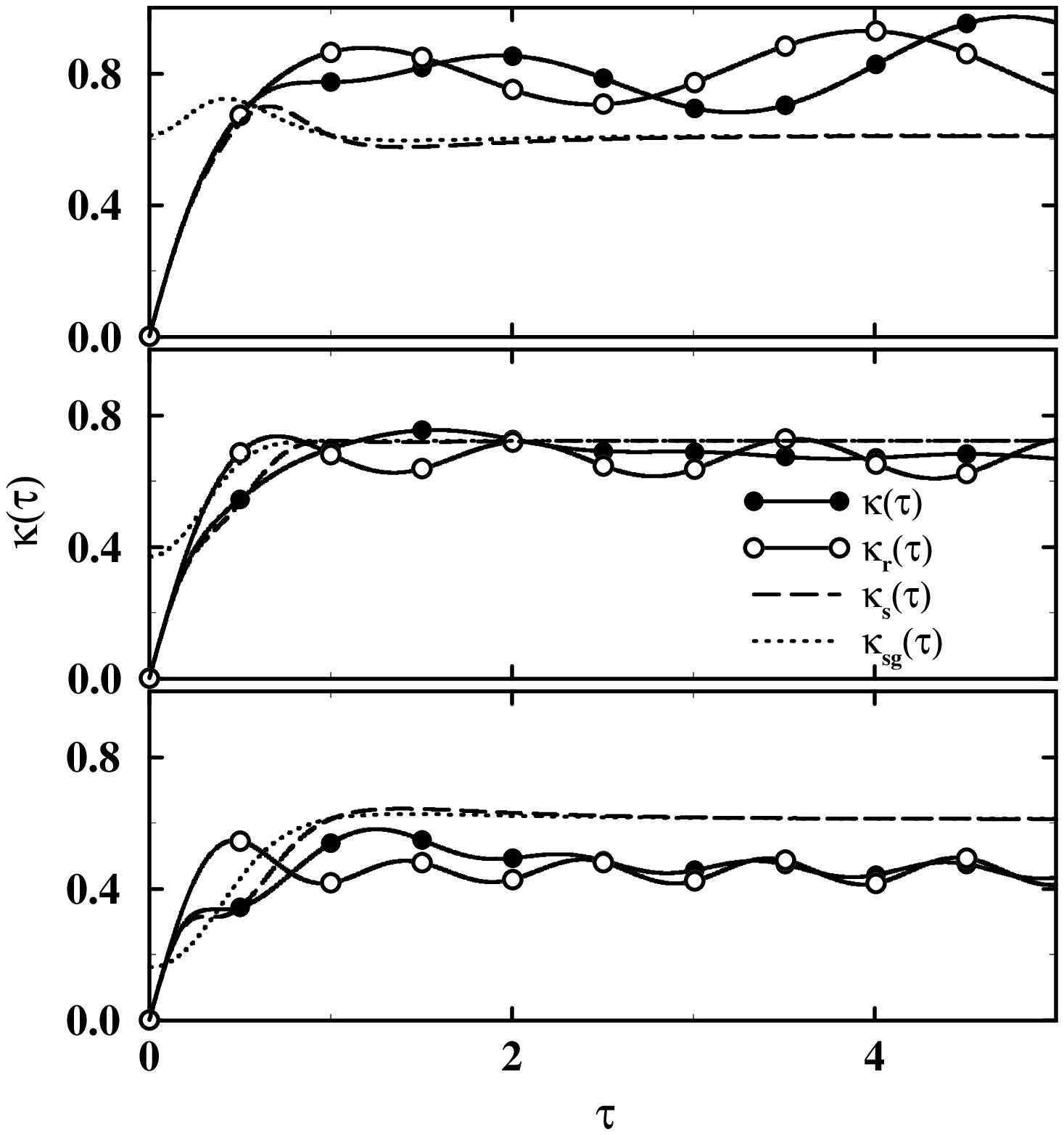}
\caption{Dimensionless nonequilibrium reaction rate $\kappa (\tau)$, Eq. (\ref{eq:kappa}), and analogous quantities for $k_r(t)$, $k_s(t)$, and $k_{sg}(t)$.
$\alpha_{_D}=\alpha_{_A}=1$ and temperature is zero.  The top panel corresponds to $\delte=\lambda_T-\lambda_D$, the middle panel to $\delte=\lambda_T$, and the bottom panel to $\delte=\lambda_T+\lambda_D$.}
\end{figure}

\begin{figure}
\epsfxsize=6 in
\epsffile{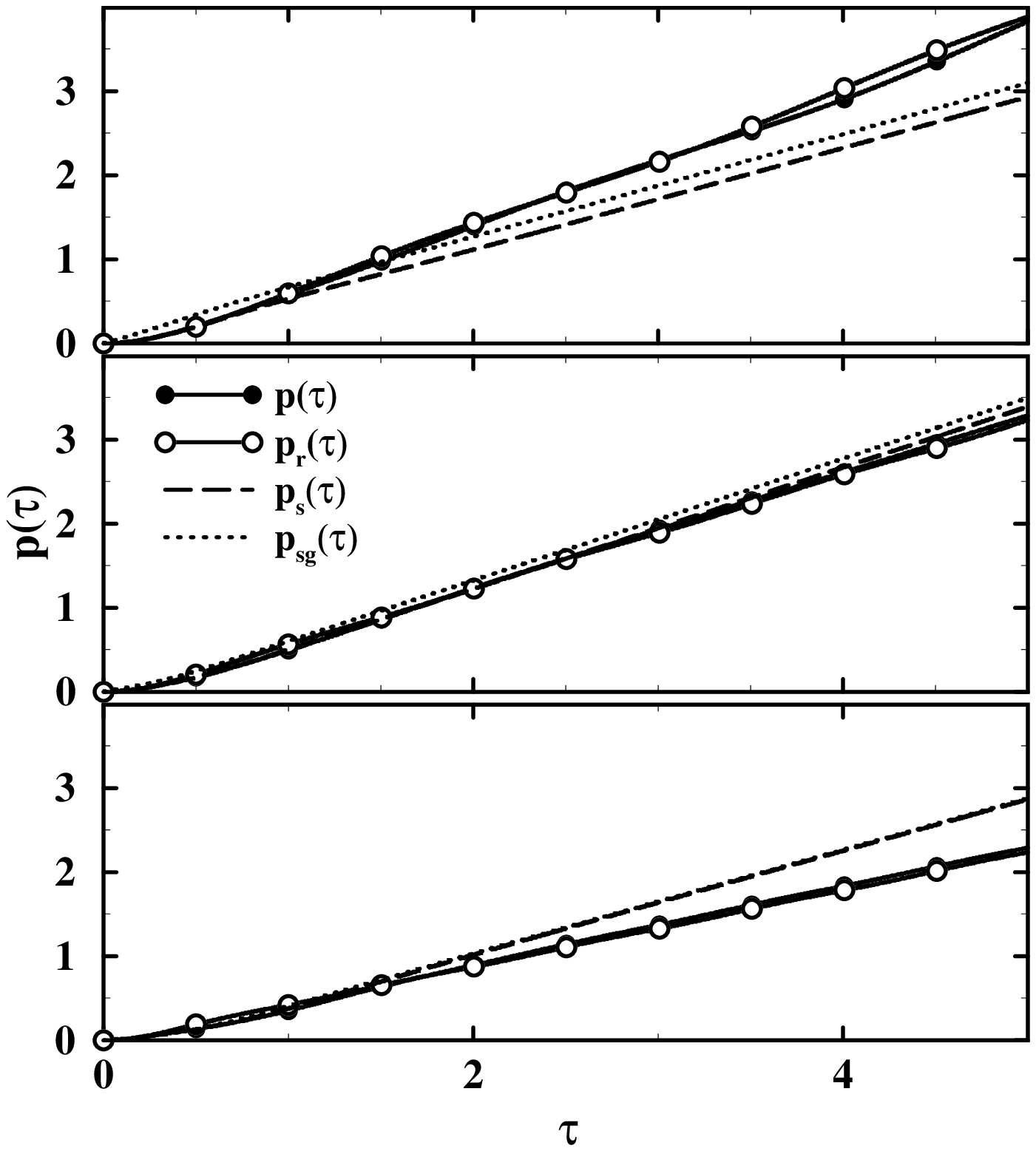}
\caption{Scaled population $p(\tau)$, Eq. (\ref{eq:pop}), and analogous quantities
for  $k_r(t)$, $k_s(t)$, and $k_{sg}(t)$. 
$\alpha_{_D}=\alpha_{_A}=1$ and temperature is zero.  The top panel corresponds to $\delte=\lambda_T-\lambda_D$, the middle panel to $\delte=\lambda_T$, and the bottom panel to $\delte=\lambda_T+\lambda_D$.}
\end{figure}

\begin{figure}
\epsfxsize=6 in
\epsffile{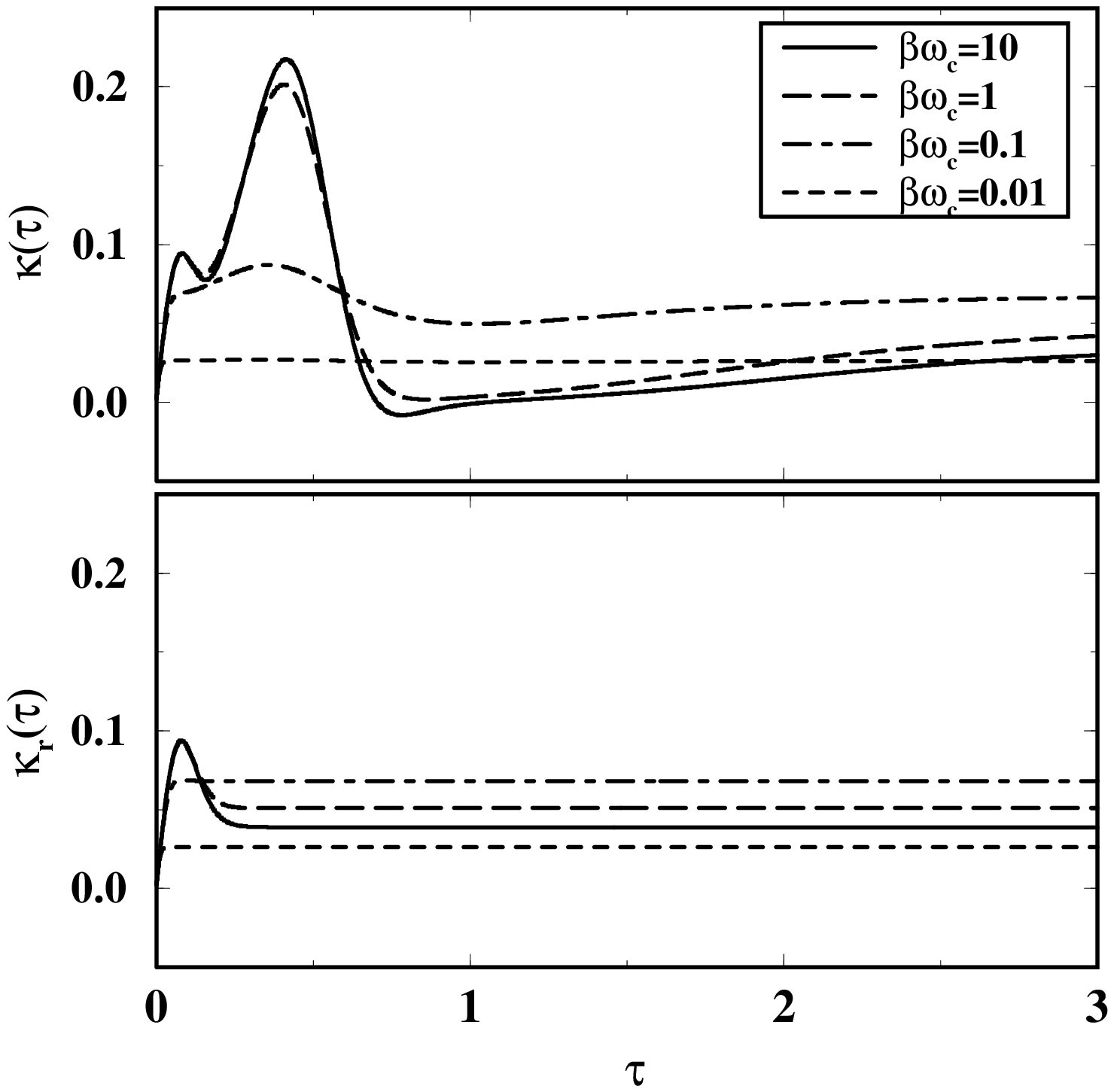}
\caption{Temperature dependence of $\kappa (\tau)$ (upper panel) and $\kappa_r (\tau)$ (lower panel) for $\alpha_{_D}=\alpha_{_A}=10$ and $\delte=\lambda_T-\lambda_D$.}
\end{figure}

\begin{figure}
\epsfxsize=6 in
\epsffile{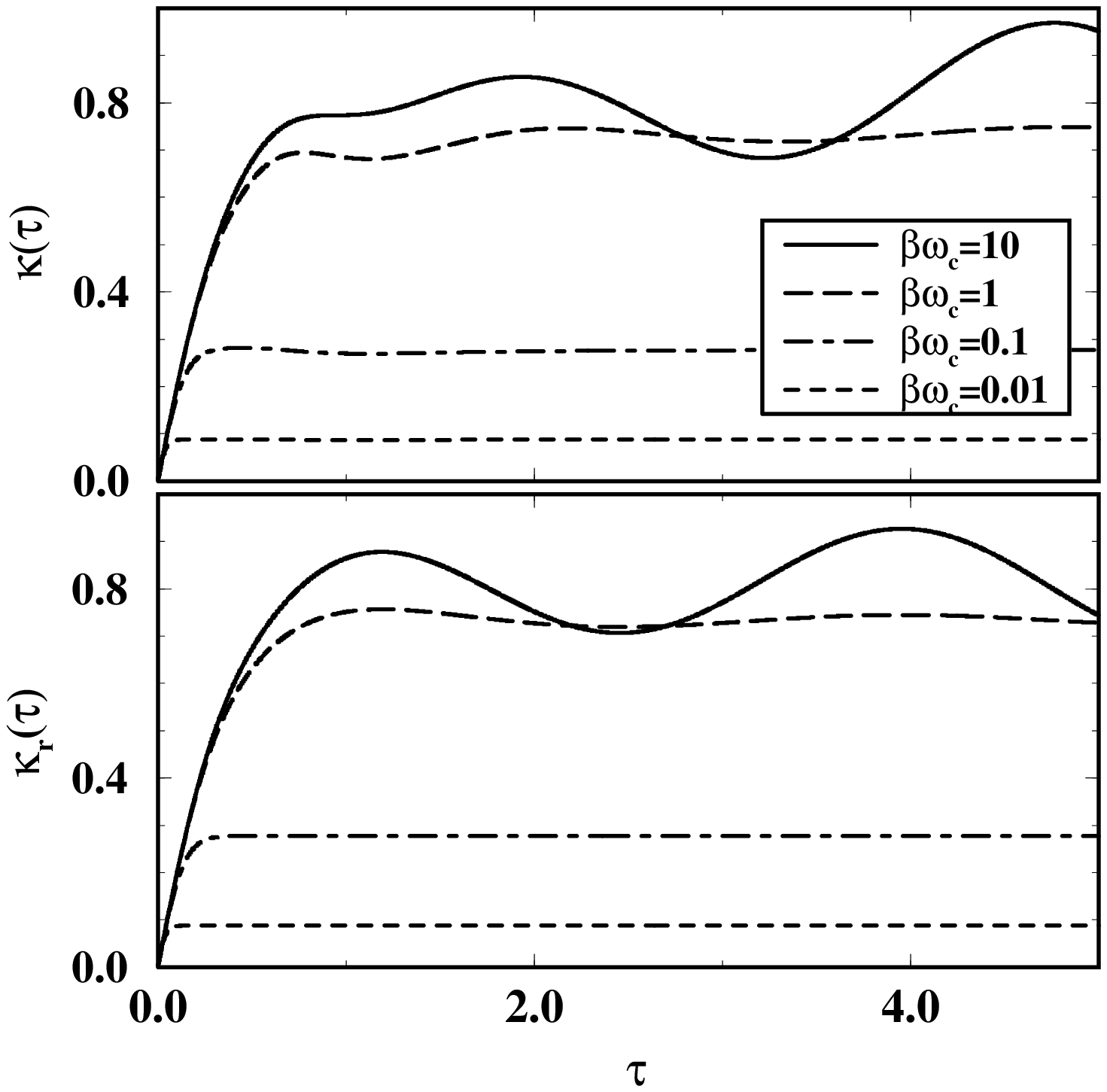}
\caption{Temperature dependence of $\kappa (\tau)$ (upper panel) and 
$\kappa_r$ (lower panel) for  $\alpha_{_D}=\alpha_{_A}=1$ and 
$\delte=\lambda_T-\lambda_D$.}
\end{figure}

\end{document}